\documentstyle[aps]{revtex}
\begin{document}
\author{Kaige Wang}
\address{CCAST (World Laboratory), P. O. Box 8730, Beijing 100080, and \\
Department of Physics, Beijing Normal University, Beijing 100875, China}
\title{Coalescence and Anti-Coalescence Interference of Two-Photon Wavepacket in a
Beam Splitter }
\maketitle

\begin{abstract}
We study a general theory on the interference of two-photon wavepacket in a
beam splitter (BS). We find that the perfect coalescence interference
requires a symmetric spectrum of two-photon wavepacket which can be
entangled or un-entangled. Furthermore, we introduce a two-photon wavepacket
with anti-symmetric spectrum, which is related with photon entanglement and
shows a perfect anti-coalescence effect. The theory present uniform and
complete explanation to two-photon interference.

PACS number(s): 42.50.Dv, 03.65.Bz, 42.50.Ar
\end{abstract}

The photon interference is one of the debated topics in the history of
quantum physics. In classical regime, interference arises from coherent
superposition of waves. In quantum systems, interference may involve in two
conflictive aspects, wave-like and particle-like, and it therefore contains
deeper and comprehensive content. In principal, quantum interference
originates from indistinguishability. When two identical particles,
physically separated, meat together, they become indistinguishable and
therefore the interference occurs. The first experimental observations of
two-photon interference in a beam splitter (BS) were reported in 80 decades
of last century \cite{mandel}\cite{shih1}. The key of experiment is to
prepare two separated and identical photons, with the same frequency and
polarization. This can be done by optical parametric down-conversion (OPDC)
of type I in a crystal, in which a pair of photons, signal and idle, are
produced. In the degenerate case, a pair of identical signal and idle
photons incident into two ports of a 50/50 BS, no coincidence count is found
at output ports. This effect is called the photon coalescence interference.
It is well known that, a pair of photons in OPDC are entangled and shown
peculiar quantum phenomena \cite{shih} such as ghost interference \cite
{shih2}, quantum eraser \cite{chiao}, quantum optical lithography \cite
{shih3} and quantum teleportation, \cite{bou}\cite{shih4}etc. In an
elaborate design on two-photon interference experiment, contributed by
Shih's group \cite{shih}\cite{shih5}, the individual signal photon and the
idle photon are arranged out of their coherent range, but the interference
is still observed. So they illustrated that the photon entanglement plays an
essential role in interference. A natural question is, can two un-entangled
identical photons show a coalescence interference?

Just recently, Santori et al \cite{san} has demonstrated in their
experiment, that indistinguishable single-photon pulses emitted by a
semiconductor quantum dot show a coalescence interference in a 50/50 beam
splitter. If there is no evidence to show these single-photon pulses are
entangled, one should believe the positive answer to the question.

Theoretically, it has known that two identical photons in an ideal
monochromatic frequency show a coalescence interference in a 50/50 BS,
whereas for two photons in different frequency, there is no interference, i.
e. 50\% coincidence probability is measured. In any realistic regime, a
finite frequency range for beams must be taken into account. If two photons
are distributed in a wider frequency range, can they show a perfect
coalescence interference? Moreover, is the photon entanglement necessary in
this effect, and what role does it play? To answer these questions, in this
paper, we present a general theory for coalescence interference of a two
photon wavepacket. We find that the perfect coalescence interference
requires a symmetric spectrum for two-photon wavepacket and photon
entanglement is not necessary. As for the un-entangled case, this condition
means that two separate single-photon wavepackets must be identical, no
matter how wider of their spectra. Furthermore, if two-photon wavepacket has
an anti-symmetric spectrum, it exhibits a new interference effect, the
anti-coalescence, in which the input state is trapped in BS. The
anti-coalescence is relevant to photon entanglement. The theory contributes
uniform and complete understanding to the previous experiments of two-photon
interference.

For a lossless BS, the transform between the input and output field
operators satisfies\cite{cam} 
\begin{equation}
\left( 
\begin{array}{c}
b_1 \\ 
b_2
\end{array}
\right) =S(\theta ,\phi _\tau ,\phi _\rho )\left( 
\begin{array}{c}
a_1 \\ 
a_2
\end{array}
\right) =\left( 
\begin{array}{cc}
e^{i\phi _\tau }\cos \theta & e^{i\phi _\rho }\sin \theta \\ 
-e^{-i\phi _\rho }\sin \theta & e^{-i\phi _\tau }\cos \theta
\end{array}
\right) \left( 
\begin{array}{c}
a_1 \\ 
a_2
\end{array}
\right) ,  \label{1}
\end{equation}
where $a_i$ and $b_i$ are the field annihilation operators for the input and
output ports, respectively. The subscripts $i=1,2$ symbolize the ports in
different direction. It is necessary to note that the transform is valid
only when two input beams are in a same mode. Otherwise, one of the input
ports must be in a vacuum state for this mode. The transform of wavevectors
in the S-picture corresponding to the operator transform (\ref{1}) is
written as $|\Psi \rangle _{out}=U|\Psi \rangle _{in}$ where $U$ was shown
in Ref. \cite{cam}. In this paper, we use a simpler method to evaluate
output wavevector. It is a duplicate of the similar method in a dynamic
quantum system \cite{kaige}, if the evolution of operators has been known in
the H-picture, one may obtain the state evolution in the S-picture without
solving the Sch\"{o}dinger equation. Assume the input state can be written
as $|\Psi \rangle _{in}=f(a_1^{\dagger },a_2^{\dagger })|0\rangle $, the
output state is obtained as 
\begin{equation}
|\Psi \rangle _{out}=Uf(a_1^{\dagger },a_2^{\dagger })|0\rangle
=Uf(a_1^{\dagger },a_2^{\dagger })U^{-1}U|0\rangle =Uf(a_1^{\dagger
},a_2^{\dagger })U^{-1}|0\rangle =f(Ua_1^{\dagger }U^{-1},Ua_2^{\dagger
}U^{-1})|0\rangle =f(\overline{b}_1^{\dagger },\overline{b}_2^{\dagger
})|0\rangle ,  \label{2}
\end{equation}
where $\overline{b}_i^{\dagger }\equiv Ua_i^{\dagger }U^{-1}$. Because $%
b_i=U^{-1}a_iU$ is known by Eq. (\ref{1}), one may obtain its inverse
transform as 
\begin{equation}
\left( 
\begin{array}{c}
\overline{b}_1 \\ 
\overline{b}_2
\end{array}
\right) =S^{-1}(\theta ,\phi _\tau ,\phi _\rho )\left( 
\begin{array}{c}
a_1 \\ 
a_2
\end{array}
\right) =S(-\theta ,-\phi _\tau ,\phi _\rho )\left( 
\begin{array}{c}
a_1 \\ 
a_2
\end{array}
\right) .  \label{3}
\end{equation}

A two-photon wavepacket is expressed as 
\begin{equation}
|\Phi _2\rangle _{in}=\sum_{\omega _1,\omega _2}C(\omega _1,\omega
_2)a_1^{\dagger }(\omega _1)a_2^{\dagger }(\omega _2)|0\rangle ,  \label{4}
\end{equation}
where the modes $a_1$ and $a_2$ have the same polarization, but are
physically separated for entering BS. $C(\omega _1,\omega _2)$ denotes a
spectrum of two-photon wavepacket. To see it, let the field operator for
mode $i$ being $E_i^{(+)}(z,t)\equiv \sum_\omega a_i(\omega )\exp [i\omega
(z/c-t)]$, we may calculate the two-photon wavepacket for state (\ref{4}) as 
\begin{equation}
\langle 0|E_1^{(+)}(z_1,t_1)E_2^{(+)}(z_2,t_2)|\Phi _2\rangle =\sum_{\omega
_1,\omega _2}C(\omega _1,\omega _2)e^{i\omega _1(z_1/c-t_1)}e^{i\omega
_2(z_2/c-t_2)}\equiv \widetilde{C}(z_1/c-t_1,z_2/c-t_2).  \label{5}
\end{equation}
If $C(\omega _1,\omega _2)=C_1(\omega _1)C_2(\omega _2)$, the two-photon
wavepacket consists of two separable single-photon wavepackets 
\begin{equation}
\langle 0|E_1^{(+)}(z_1,t_1)E_2^{(+)}(z_2,t_2)|\Phi _2\rangle =\sum_{\omega
_1,\omega _2}C_1(\omega _1)C_2(\omega _2)e^{i\omega _1(z_1/c-t_1)}e^{i\omega
_2(z_2/c-t_2)}\equiv \widetilde{C}_1(z_1/c-t_1)\widetilde{C}_2(z_2/c-t_2).
\label{6}
\end{equation}
Thus, the two-photon state $|\Phi _2\rangle $ is un-entangled.

By using Eqs. (\ref{2}) and (\ref{3}), for the input state $|\Phi _2\rangle
_{in},$ the output state of BS is obtained as 
\begin{eqnarray}
|\Phi _2\rangle _{out} &=&\sum_{\omega _1,\omega _2}C(\omega _1,\omega _2)%
\overline{b}_1^{\dagger }(\omega _1)\overline{b}_2^{\dagger }(\omega
_2)|0\rangle   \label{7} \\
&=&\sum_{\omega _1,\omega _2}C(\omega _1,\omega _2)[a_1^{\dagger }(\omega
_1)e^{i\phi _\tau }\cos \theta -a_2^{\dagger }(\omega _1)e^{-i\phi _\rho
}\sin \theta ][a_1^{\dagger }(\omega _2)e^{i\phi _\rho }\sin \theta
+a_2^{\dagger }(\omega _2)e^{-i\phi _\tau }\cos \theta ]|0\rangle   \nonumber
\\
&=&\sum_{\omega _1,\omega _2}C(\omega _1,\omega _2)\{[a_1^{\dagger }(\omega
_1)a_1^{\dagger }(\omega _2)e^{i\phi }-a_2^{\dagger }(\omega _1)a_2^{\dagger
}(\omega _2)e^{-i\phi }]\cos \theta \sin \theta   \nonumber \\
&&+[a_1^{\dagger }(\omega _1)a_2^{\dagger }(\omega _2)\cos ^2\theta
-a_2^{\dagger }(\omega _1)a_1^{\dagger }(\omega _2)\sin ^2\theta
]\}|0\rangle ,  \nonumber
\end{eqnarray}
where $\phi =\phi _\tau +\phi _\rho $. It shows the output consists of four
states which are illustrate in the first line of Fig. 1. Because the
summation is taken over the whole frequency space, there are also other four
states, by exchanging $\omega _1$ and $\omega _2,$ as shown in the second
line of Fig. 1. The states in each column of Fig. 1 are indistinguishable
and should put together. Therefore, we take $\sum_{\omega _1,\omega
_2}=\sum_{\omega _1<\omega _2}+\sum_{\omega _1=\omega _2}+\sum_{\omega
_1>\omega _2}$, and then exchange the variables $\omega _1$ and $\omega _2$
in the last summation, Eq. (\ref{7}) is written as 
\begin{eqnarray}
|\Phi _2\rangle _{out} &=&\sum_{\omega _1<\omega _2}\{[C(\omega _1,\omega
_2)+C(\omega _2,\omega _1)][a_1^{\dagger }(\omega _1)a_1^{\dagger }(\omega
_2)e^{i\phi }-a_2^{\dagger }(\omega _1)a_2^{\dagger }(\omega _2)e^{-i\phi
}]\cos \theta \sin \theta   \label{8} \\
&&+[C(\omega _1,\omega _2)\cos ^2\theta -C(\omega _2,\omega _1)\sin ^2\theta
]a_1^{\dagger }(\omega _1)a_2^{\dagger }(\omega _2)  \nonumber \\
&&+[C(\omega _2,\omega _1)\cos ^2\theta -C(\omega _1,\omega _2)\sin ^2\theta
]a_1^{\dagger }(\omega _2)a_2^{\dagger }(\omega _1)\}|0\rangle   \nonumber \\
&&+\sum_\omega C(\omega ,\omega )\{[(a_1^{\dagger }(\omega ))^2e^{i\phi
}-(a_2^{\dagger }(\omega ))^2e^{-i\phi }]\cos \theta \sin \theta +(\cos
^2\theta -\sin ^2\theta )a_1^{\dagger }(\omega )a_2^{\dagger }(\omega
)\}|0\rangle .  \nonumber
\end{eqnarray}
For a 50/50 BS, Eq. (\ref{8}) is reduced as 
\begin{eqnarray}
|\Phi _2\rangle _{out} &=&(1/\sqrt{2})\sum_{\omega _1<\omega _2}\{[C(\omega
_1,\omega _2)+C(\omega _2,\omega _1)][|1(\omega _1),1(\omega _2);0\rangle
e^{i\phi }-|0;1(\omega _1),1(\omega _2)\rangle e^{-i\phi }]/\sqrt{2}
\label{9} \\
&&+[C(\omega _1,\omega _2)-C(\omega _2,\omega _1)][|1(\omega _1);1(\omega
_2)\rangle -|1(\omega _2);1(\omega _1)\rangle ]/\sqrt{2}\}  \nonumber \\
&&+\sum_\omega C(\omega ,\omega )[|2(\omega );0\rangle e^{i\phi
}-|0;2(\omega )\rangle e^{-i\phi }]/\sqrt{2}.  \nonumber
\end{eqnarray}
In the braket, the left part of the semicolon describes the photon state in
one output port, whereas the right part, the photon state in another output
port. $1(\omega )$ and $2(\omega )$ define respectively one-photon and
two-photon Fock states with a frequency $\omega $. In Eq. (\ref{9}), the
first and last terms describe photon coalescence, whereas the second term
describes photon anti-coalescence resulting in ''click-click'' in a
coincidence measurement. Two manners of photon interference depend on the
symmetry of two-photon spectrum. If a two-photon wavepacket has a symmetric
spectrum $C(\omega _1,\omega _2)=C(\omega _2,\omega _1)$ in whole frequency
space, the second term vanishes and a perfect coalescence interference
occurs. A symmetric spectrum can be acquired in both entangled and
un-entangled two-photon wavepacket. In the un-entangled case, the symmetric
spectrum $C_1(\omega _1)C_2(\omega _2)=C_2(\omega _1)C_1(\omega _2)$ implies 
$C_1(\omega _1)/C_2(\omega _1)$ independent of frequency, and hence, the two
single-photon wavepackets must be identical $C_1(\omega )=C_2(\omega )$.

Another manner of two-photon interference is just opposite of coalescence
interference: two photons never go together and cause a definite click-click
coincidence. We call it the anti-coalescence interference. It happens when
two-photon spectrum is anti-symmetric $C(\omega _1,\omega _2)=-C(\omega
_2,\omega _1)$ in whole frequency space, and, obviously, it also satisfies $%
C(\omega ,\omega )\equiv 0$. In this case, photon coalescence cancels
exactly and the output state (\ref{9}) is reduced to 
\begin{eqnarray}
|\Phi _2\rangle _{out} &=&\sum_{\omega _1<\omega _2}C(\omega _1,\omega
_2)[a_1^{\dagger }(\omega _1)a_2^{\dagger }(\omega _2)-a_1^{\dagger }(\omega
_2)a_2^{\dagger }(\omega _1)]|0\rangle   \label{9p} \\
&=&\sum_{\omega _1<\omega _2}C(\omega _1,\omega _2)a_1^{\dagger }(\omega
_1)a_2^{\dagger }(\omega _2)|0\rangle -\sum_{\omega _2<\omega _1}C(\omega
_2,\omega _1)a_1^{\dagger }(\omega _1)a_2^{\dagger }(\omega _2)]|0\rangle  
\nonumber \\
&=&\sum_{\omega _1,\omega _2}C(\omega _1,\omega _2)a_1^{\dagger }(\omega
_1)a_2^{\dagger }(\omega _2)|0\rangle =|\Phi _2\rangle _{in}.  \nonumber
\end{eqnarray}
It means that an anti-symmetric two-photon wavepacket is invariant under the
50/50 BS transform. We see again an interesting example of quantum
destructive interference --- all interactions are cancelled each with other
exactly and quantum state is trapped. This phenomenon is really surprise
since two-photon anti-symmetric spectrum has the same power spectrum as the
symmetric one. Out off phase with exchange of $\omega _1$ and $\omega _2$ in
a two-photon spectrum makes two identical single-photon wavepackets ''do not
know each other completely''. Physically, two identical photons never do
anti-coalescence interference. As a matter of fact, in a two-photon
wavepacket with an anti-symmetric spectrum, there is no any pair of
identical photons because of $C(\omega ,\omega )\equiv 0.$ Therefore,
anti-coalescence is an interference effect based on two different photons.
However, the important fact is that it is impossible to realize an
anti-symmetric wavepacket for any un-entangled two-photon state. Since, in
the anti-symmetric case, $C_1(\omega _1)C_2(\omega _2)=-C_2(\omega
_1)C_1(\omega _2)$ should be valid also for $\omega _1=\omega _2=\omega $,
it gives $C_1(\omega )C_2(\omega )=0$. So the anti-coalescence effect relies
definitely on photon entanglement. A well-known example of anti-symmetric
two-photon state is the anti-symmetric Bell state, which can be expressed as 
$C(\omega _1,\omega _2)=[\delta (\omega _1-\Omega _1)\delta (\omega
_2-\Omega _2)-\delta (\omega _1-\Omega _2)\delta (\omega _2-\Omega _1)]/%
\sqrt{2}$. Because of the trapping effect, any two-photon state with an
anti-symmetric spectrum is the eigenstate in a 50/50 BS transform with the
unity eigenvalue. This feature has been applied in teleportation scheme \cite
{bou}. The anti-coalescence effect was observed in experiments \cite{chiao}%
\cite{shih5}, in which, instead of dip, an peak appears in a coincidence
measurement.

In a general case, we may evaluate the click-click probability as 
\begin{equation}
P=(1/2)\sum_{\omega _1<\omega _2}|C(\omega _1,\omega _2)-C(\omega _2,\omega
_1)|^2=(1/2)\int_{-\infty }^\infty d\omega _1\int_{-\omega _1}^\infty
d\omega _2|C(\omega _1,\omega _2)-C(\omega _2,\omega _1)|^2.  \label{10}
\end{equation}
$P<1/2$ ($P>1/2$) denotes a coalescence (anti-coalescence) interference.
Because the integrand is a symmetric function with exchange $\omega _1$ and $%
\omega _2$, and vanishes at $\omega _1=\omega _2$, the above integration can
be written as 
\begin{equation}
P=(1/4)\int_{-\infty }^\infty d\omega _1\int_{-\infty }^\infty d\omega
_2|C(\omega _1,\omega _2)-C(\omega _2,\omega _1)|^2.  \label{11}
\end{equation}

Now we discuss coherent range of coalescence interference. Assume a spectrum 
$C(\omega _1,\omega _2)$ being symmetric, we add some phase shifts on both
beams. According to Eq. (\ref{6}), the new spectrum is then $C(\omega
_1,\omega _2)e^{i(\omega _1z_1/c+\omega _2z_2/c)}$ which is asymmetric if $%
z_1\neq z_2$. The click-click probability is obtained as 
\begin{eqnarray}
P &=&(1/2)\int_{-\infty }^\infty d\omega _1\int_{-\infty }^\infty d\omega
_2|C(\omega _1,\omega _2)|^2\{1-\cos [(\omega _1-\omega _2)\Delta z/c]\}
\label{12} \\
&=&(1/2)\{1-\int_{-\infty }^\infty d\omega _1\int_{-\infty }^\infty d\omega
_2|C(\omega _1,\omega _2)|^2\cos [(\omega _1-\omega _2)\Delta z/c]\}, 
\nonumber
\end{eqnarray}
where $\Delta z=z_1-z_2$. Consider a two-photon wavepacket with a symmetric
spectrum as 
\begin{equation}
C(\omega _1,\omega _2)=g(\omega _1+\omega _2)e^{-[(\omega _1-\Omega
)^2+(\omega _2-\Omega )^2]/(2\sigma ^2)},  \label{13}
\end{equation}
where the exponential function describes two single-photon spectra with the
same bandwidth $\sigma $, and $\Omega $ is their identical centre frequency. 
$g(\omega _1+\omega _2)$ may describe two-photon entanglement. For example,
it can be a Gaussian type in OPDC of type I \cite{rubin}. If $g(\omega
_1+\omega _2)$ is a constant, Eq. (\ref{13}) reduces to an un-entangled
two-photon wavepacket which spectrum has been shown in Fig. 2a. Substituting
Eq. (\ref{13}) into Eq. (\ref{12}), one obtains 
\begin{equation}
P=(1/2)(1-e^{-\frac 12(\sigma \Delta z/c)^2}).  \label{14}
\end{equation}
This equation displays the famous dip observed in the experiments of
two-photon interference. A perfect coalescence exists at $\Delta z=0$, no
matter how wider the spectrum is. The width of the dip $w=c/\sigma $
indicates a coherent length of single-photon. Note that the maximum
coincidence probability in Eq. (\ref{14}) is one half showing no
interference. However, this is an uniform result for both entangled and
un-entangled two-photon wavepacket, since the coincidence probability does
not depend on idiographic form of function $g$. This conforms the previous
prediction --- the entanglement is not a necessary condition for coalescence
interference. The key point is that one needs to create a pair of identical
single-photon wavepacket, the same centre frequency and the same bandwidth.

To show an anti-coalescence effect, we study again the previous model in
Ref. \ref{shih5} by this theory. In OPDC of type I, it has $g(\omega
_1+\omega _2)\sim e^{-(\omega _1+\omega _2-2\Omega )^2/(2\sigma _p^2)}$
where $\sigma _p$ is the bandwidth of the pump beam \cite{rubin}. In Ref. 
\cite{shih5}, the signal beam was managed in both a short and a long paths,
so that the state entering BS is a coherent superposition of the states for
these two paths $L_s$ and $L_l$%
\begin{eqnarray}
C(\omega _1,\omega _2) &=&Ae^{-(\omega _1+\omega _2-2\Omega )^2/(2\sigma
_p^2)}e^{-[(\omega _1-\Omega )^2+(\omega _2-\Omega )^2]/(2\sigma
^2)}(1/2)\{e^{i(\omega _1L_s+\omega _2z_2)/c}+e^{i(\omega _1L_l+\omega
_2z_2)/c}\}  \label{16} \\
&=&Ae^{-(\omega _1+\omega _2-2\Omega )^2/(2\sigma _p^2)}e^{-[(\omega
_1-\Omega )^2+(\omega _2-\Omega )^2]/(2\sigma ^2)}e^{i(\omega _1z_1+\omega
_2z_2)/c}\cos (\omega _1\Delta L/c),  \nonumber
\end{eqnarray}
where $\Delta L=(L_l-L_s)/2$; $z_1=(L_l+L_s)/2$ is the average path of the
signal beam, and $z_2$ is the path of the idle beam. We calculate the exact
coincidence probability for two-photon spectrum (\ref{16}) and obtain 
\begin{eqnarray}
P &=&\frac 12\{1-\frac 1{2B}[\cos (\frac{4\pi \Delta L}\lambda )e^{-\frac 12[%
\frac{\beta ^2}{2+\beta ^2}\Delta L^2+\Delta z^2](\frac \sigma c)^2}+\frac 12%
e^{-\frac 12(\Delta L+\Delta z)^2(\frac \sigma c)^2}+\frac 12e^{-\frac 12%
(\Delta L-\Delta z)^2(\frac \sigma c)^2}]\},  \label{17} \\
B &=&\frac 12[1+\cos (\frac{4\pi \Delta L}\lambda )e^{-\frac 14\frac{1+\beta
^2}{2+\beta ^2}\Delta L^2(\frac \sigma c)^2}],  \nonumber
\end{eqnarray}
where $\beta \equiv \sigma _p/\sigma ,$ and $\lambda =2\pi c/\Omega $ is the
wavelength of the signal beam. In the case the path difference is much
larger than the coherent length of the single-photon, $\Delta L>>c/\sigma $,
one has $B\rightarrow 1/2$. However, by taking into account $\beta <<1$, Eq.
(\ref{17}) is reduced as 
\begin{equation}
P=\frac 12\{1-\cos (\frac{4\pi \Delta L}\lambda )e^{-\frac 12\Delta z^2(%
\frac \sigma c)^2}-\frac 12e^{-\frac 12(\Delta L+\Delta z)^2(\frac \sigma c%
)^2}-\frac 12e^{-\frac 12(\Delta L-\Delta z)^2(\frac \sigma c)^2}]\}.
\label{19}
\end{equation}
This is exactly the result as in Ref. \cite{shih5}. If one sets $4\Delta
L/\lambda =1,3,5,\cdots $, one obtains the peak ($P\rightarrow 1)$ at $%
\Delta z=0$.

We assume, in an ideal case, the pump beam has an infinitesimal bandwidth,
so that the two-photon wavepacket can be written as 
\begin{eqnarray}
C(\omega _1,\omega _2) &=&A\delta (\omega _1+\omega _2-2\Omega )e^{-[(\omega
_1-\Omega )^2+(\omega _2-\Omega )^2]/(2\sigma ^2)}\cos (\omega _1\Delta L/c)
\label{20} \\
&=&A\delta (\nu _1+\nu _2)e^{-(\nu _1^2+\nu _2^2)/(2\sigma ^2)}%
%TCIMACRO{
%\QATOPD\{ . {\cos (\nu _1\Delta L/c)\qquad \text{for }4\Delta L/\lambda =0,2,4,\cdots }{\sin (\nu _1\Delta L/c)\qquad \text{for }4\Delta L/\lambda =1,3,5,\cdots }}
%BeginExpansion
{\cos (\nu _1\Delta L/c)\qquad \text{for }4\Delta L/\lambda =0,2,4,\cdots  \atopwithdelims\{. \sin (\nu _1\Delta L/c)\qquad \text{for }4\Delta L/\lambda =1,3,5,\cdots }
%EndExpansion
,  \nonumber
\end{eqnarray}
where $\nu _i=\omega _i-\Omega .$ Because $\delta (\nu _1+\nu _2)$ implies $%
\nu _1=-\nu _2,$ the symmetric and the anti-symmetric spectra correspond to
the forms of cosine and sine, as shown in Fig 2b and 2c, respectively.
Obviously, there is no degenerate photons ($\nu _1=\nu _2=0$) in the
anti-symmetric wavepacket.

In conclusion, we demonstrate that, in a general theory, the symmetry of
two-photon spectrum dominates coalescence and anti-coalescence
interferences. The theory clarifies debatable aspect in two photon
interference. The photon entanglement is not a precondition for coalescence
interference and two un-entangled single-photon wavepackets can interfere in
a BS provided their spectra are identical. The new interference effects,
anti-coalescence and trapped two-photon wavepacket, are brought forward and
noteworthy owing to their relation with photon entanglement. Coalescence and
anti-coalescence effects can be understood as indistinguishability of two
two-photon states and their constructive or destructive coherence. The
couple of two-photon states, forming the symmetric (anti-symmetric)
Bell-type combination in the input, contributes to coalescence
(anti-coalescence) interference. However, in the degenerate case, the
coalescence interference of two identical photons is the result of
cancellation of two anti-coalescence two-photon states. In this sense,
two-photon interference does not conflict with Dirac's famous statement
''...photon... only interferes with itself. Interference between two
different photons never occurs''. What leaves over is the amazing quantum
nature: in two-photon wavepacket with symmetry, both entangled and
un-entangled, the two-photon state $a_1^{\dagger }(\omega _1)a_2^{\dagger
}(\omega _2)|0\rangle $ coalesces with $a_1^{\dagger }(\omega
_2)a_2^{\dagger }(\omega _1)|0\rangle $ automatically, resulting in perfect
interference.

This research is founded by the National Fundamental Research Program of
China with No. 2001CB309310, and the National Natural Science Foundation of
China, Project Nos. 60278021 and 10074008.

\bigskip\ 

Captions of figures:

Fig. 1 Scheme of two-photon states produced in beam splitter.

Fig. 2 Two-photon wavepackets for (a) un-entangled state, (b) entangled
state with a symmetric spectrum and (c) entangled state with an
anti-symmetric spectrum. The peaks correspond to the centre carrier
frequency $\Omega $.

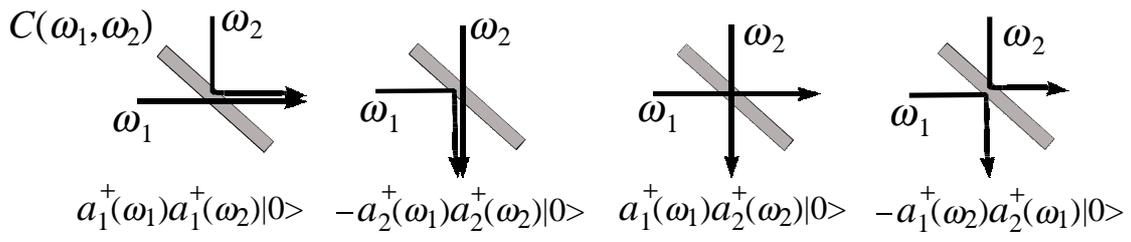
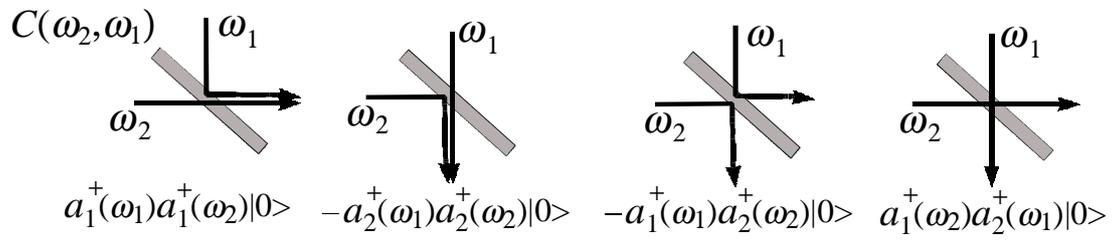
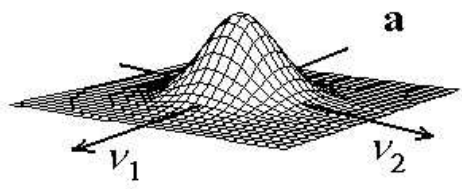
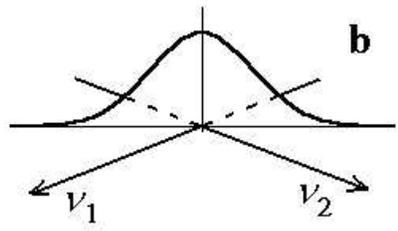
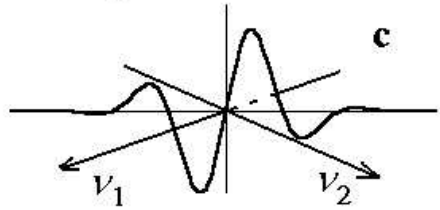
\end{document}